\documentstyle[12pt]{article}
\newcommand{\be}{\begin{equation}}
\newcommand{\ee}{\end{equation}}
\newcommand{\bea}{\begin{eqnarray}}
\newcommand{\eea}{\end{eqnarray}}
\newcommand{\ep}{\varepsilon}

\newcommand{\Gm}{\Gamma}
\newcommand{\gm}{\gamma}
\newcommand{\om}{\omega}
\newcommand{\lm}{\lambda}
\newcommand{\oR}{\overline{R}}
\newcommand{\uQ}{\underline{Q}}
\newcommand{\uq}{\underline{q}}
\newcommand{\uM}{\underline{M}}
\newcommand{\um}{\underline{m}}
\newcommand{\uk}{\underline{k}}
\newcommand{\unu}{\underline{\nu}}
\newcommand{\dd}{\mbox{d}}
\newcommand{\nn}{\nonumber}

\newcommand{\Li}[2]{{\mbox{Li}}_{#1}\left(#2\right)}

\begin{document}
\parindent=1.5pc
\setlength {\unitlength}{1mm}
\begin{titlepage}
\rightline{MPI--PhT/94--82}
\rightline{hep-th/9412063}
\rightline{November 1994}

\begin{center}
{{\bf
Asymptotic Expansions in Momenta and Masses \\
and Calculation of Feynman Diagrams\footnote{Talk given at the workshop
``Advances in non-perturbative and perturbative techniques'',
Ringberg Castle, November 13--19, 1994.}}\\
\vglue 5pt
\vglue 1.0cm
{ {\large V.A. Smirnov}\footnote{
Permanent address: Nuclear Physics Institute of
Moscow State University,
Moscow 119899, Russia;
E-mail: smirnov@compnet.msu.su } }\\
\baselineskip=14pt
\vspace{2mm}
{\it Max-Planck-Institut f\"ur Physik, Werner-Heisenberg-Institut}\\
{\it F\"ohringer  Ring 6, 80805 Munich, Germany}\\
\vglue 0.8cm
{Abstract}}
\end{center}
\vglue 0.3cm
{\rightskip=3pc
 \leftskip=3pc
 \tenrm\baselineskip=12pt
\noindent
General results on asymptotic
expansions of Feynman diagrams
in momenta and/or masses are reviewed. It is shown how they
are applied for calculation of massive diagrams.
\vglue 0.8cm}
\end{titlepage}

{\bf 1.} {\em Earlier results.}

\vspace{1mm}

The problem of asymptotic\footnote{The term `asymptotic' implies
that the corresponding remainder satisfies necessary estimates and one
knows nothing about the radius of convergence. In contrast to expansions
in coupling constants which typically have zero radii of convergence,
the large mass/momentum expansions of Feynman diagrams
seem to have always non-zero
radii of convergence.}
expansions of Feynman amplitudes in momenta \linebreak
and/or masses is rather old.
A limit of large momenta and masses is characterized by a subdivision
of the set of all external momenta and internal masses of a diagram $\Gm$
into large
$\uQ \equiv \{ Q_1, \ldots Q_i, \dots \}$,
$\uM \equiv \{ M_1, \ldots M_i, \dots \}$, and small
$\uq \equiv \{ q_1, \ldots q_i, \dots \}$,
$\um \equiv \{ m_1, \ldots m_i, \dots \}$ ones.
The problem is to analyze the behavior of the corresponding Feynman
integral
$F_{\Gm} (\uQ , \uq, \uM , \um)$ in the limit
$F_{\Gm} (\uQ /\rho, \uq, \uM /\rho, \um)$ as $\rho \to 0$.
For brevity, let us denote this limit as $\uQ,\uM \to \infty$.
Let us imply that the external
momenta are not fixed at a mass shell.\footnote{The results that
are presented here hold both for Minkowski and Euclidean spaces.
For Minkowski space, it is in fact sufficient
to imply that the large external momenta are space-like. However, for
the limit of large masses when all the momenta are small, there are no
restriction on momenta. Another possible variant is to consider
Feynman diagrams as distributions in momenta.}

In 1960  Weinberg
\cite{Wei} described the leading large momentum behavior. Logarithmic
corrections were characterized in \cite{Fink}. Later it was proved that
the large momentum asymptotic expansions are always performed in powers
and logarithms of the expansion parameter \cite{Sla}. In
\cite{Zav,Ber1,Ber2,Cal,Poh,Ber3,Cas,Kos}
asymptotic expansions in various limit
of large momenta and masses were obtained. A typical result is the expansion
of the form
\be
F_{\Gm} (\uQ /\rho, \uq, \um)
\; \stackrel{\mbox{\footnotesize$\rho \to 0$}}{\mbox{\Large$\sim$}} \;
\sum_{k,l} C_{k,l}\; \rho^k \log^l \rho .
\label{F1}
\ee
However the coefficient functions in these expansions are cumbersome.
They are expressed in terms of numerous parametric integrals or in
terms of Mellin integrals.
Thus, the first of the following two natural properties of asymptotic
expansions does not hold:

({\em i}) The coefficient functions $C_{k,l}$
are expressed in a
simple way through renormalized and/or regularized Feynman amplitudes;

({\em ii}) The expansion is in powers and logarithms.

\vspace{3mm}

{\bf 2.} {\em  Results in the simplest form for the large momentum limit. }

\vspace{1mm}

To write down the asymptotic expansion with these two properties it is
worthwhile to introduce dimensional (with $d=4-2\ep$)
regularization even in
case the original diagram is ultravioletly finite. The following proposition
is valid.

In the large momentum limit,
\be
F_{\Gm} (\uQ /\rho, \uq, \um;\ep)
\; \stackrel{\mbox{\footnotesize$\uQ \to \infty$}}{\mbox{\Large$\sim$}} \;
\sum_{\gamma} F_{\Gamma / \gamma} (\uq,\um;\ep)
\circ {\cal T}_{\uq^{\gm}, \um^{\gm}} F_{\gamma} (\uQ,\uq^{\gm},
\um^{\gm};\ep),
\label{F2}
\ee
where the sum is over subgraphs $\gamma$ of $\Gamma$ such that each
$\gamma$ (a) contains all the vertices with the large external
momenta and (b) is 1PI after contraction of these vertices.
Furthermore, the operator ${\cal T}$ performs Taylor
expansion in the corresponding
set of variables; $\uq^{\gm}$ are the light external momenta of the
subgraph $\gamma$ (i.e. all its external momenta apart from
the large external momenta of $\Gamma$); $m^{\gamma}$ is a set of masses
of $\gamma$.
Finally, if $F_{\Gm / \gm}$ is a Feynman integral corresponding to the
reduced graph $\Gm / \gm$ and
${\cal P}_{\gm}$ is a polynomial in $\uq^{\gm}$, then the expression
$F_{\Gamma / \gamma} \circ {\cal P}_{\gm}$ denotes the Feynman integral
that differs from  $F_{\Gm / \gm}$ by insertion of ${\cal P}$ into the
vertex $v_{\gm}$ (this is the vertex to which the subgraph $\gm$ was
collapsed).
It is implied that the operators ${\cal T}$ act directly on the integrands of
Feynman integrals over loop momenta.

Expansion (\ref{F2}) was first written in an equivalent form in
\cite{Gor} (see also \cite{Che1,Che2,Che3}).
In a particular case of one large external momentum the expansion
was found and justified within the so-called method of glueing
\cite{Che2}.

In \cite{Piv} an expansion similar to (\ref{F2}) was presented.
Its validity was taken as a postulate.
Later the approach of \cite{Piv} resulted in the theory of
As-operation\footnote{which is characterized by its author as
`the main stream quantum field theory' and `the most important
developement since 1972' \cite{Tka1}.} \cite{Tka2}.
According to prescriptions of the As-operation,\footnote{In the case
of dimensional regularization, mathematical
proofs of results on the As-operation are really absent. Since the
basic principle of the As-operation is to apply
distribution theory, it would be necessary to explain what
are test functions in $d$ dimensions and to provide at least
one non-trivial (non-zero) example of such a test function.}
to obtain asymptotic
expansions in a given limit it is necessary (a) perform a formal
(naive) Taylor expansion in small momenta and masses, (b) study
IR divergences that are induced in this formal expansion, (c)
characterize the nature of these singularities in graph-theoretical
language, (d) find analytical structure of the singularities and
(e) construct other terms of the expansion. It should be noted that
there is no need to perform this job for each limit
and follow these multiple prescriptions: it is sufficient
to use explicit formula (\ref{F2}) and similar formulae
for other limits (see below) by writing down the corresponding
expansions from the very beginning.

Asymptotic expansion (\ref{F2}) as well asymptotic expansion in other
situations (see below) was justified in \cite{Smi1}
(see also \cite{Smi2}). There also exist another version of the proof
based on the method of glueing (respectively,
combinatorial \cite{Che3} and analytical \cite{Che2,Che4} parts).

Practically, to write down the quantity
$F_{\Gamma / \gamma} \circ {\cal T}_{\ldots} F_{\gm}$
it is necessary

(1) when choosing the loop momenta of $\Gm$, first, to choose a set of the
loop momenta of $\gm \subset \Gm$;

(2) to let large external momenta flow through $\gm$.

Thus, if
\be
F_{\Gm} (\uQ, \uq, \um;\ep) =
\int \dd k_1 \ldots \dd k_h \Pi_{\Gm} (\uQ,\uq,\uk,\um),
\label{F3}
\ee
where $\uk\equiv k_1, \ldots k_h$ is the set of the loop momenta of $\Gm$, and
$\Pi_{\Gm} \equiv \Pi_{\Gm \setminus \gm} \Pi_{\gm}$ is the product
of propagators associated with the given graph, then
\be
F_{\Gamma / \gamma} \circ
{\cal T}_{\uq^{\gm}, \um^{\gm}} F_{\gamma}
= \int \dd k_1 \ldots \dd k_h \Pi_{\Gm\setminus \gm}
{\cal T}_{\uq, \uk^{\Gm / \gm}, \um^{\gm}} \Pi_{\gamma}
\label{F4}
\ee
so that the small external momenta for the subgraph $\gm$ turn out
to be the small momenta $\uq$ of the initial graph itself as well as
the loop momenta $\uk^{\Gm / \gm}$ of the reduced graph. (Note that according
to these two rules, the loop momenta of $\Gm$ are subdivided into the loop
momenta of $\gm$ and that of $\Gm / \gm$.)

Let us now observe that both properties ({\em i}) and ({\em ii}) are satisfied
for expansion (\ref{F2}). In fact, if ${\cal T}^{(j)}$ is contribution
of the terms of the $j$-th order of the corresponding Taylor series, and
$\om(\gm)$ is the degree of divergence of $\gm$, then
\be
{\cal T}^{(j)}_{\uq, \um} F_{\gamma}  (\uQ/\rho, \uq, \um;\ep) =
\rho^{-\om(\gm) + j - \ep h(\gm)}
{\cal T}^{(j)}_{\uq, \um} F_{\gamma}  (\uQ, \uq, \um;\ep),
\ee
where $h(\gm)$ is the number of loops of $\gm$.

Since in the large momentum limit the graph itself contributes
to the sum in (\ref{F2}) there is always the term
\be
{\cal T}_{\uq, \um} F_{\Gamma} (\uQ,\uq, \um;\ep) .
\label{F6}
\ee
This is nothing but the `naive' part of the expansion, in the sense that
this contribution appears as a result of Taylor expansion of the
integrand of the initial Feynman integral with respect to the small
parameters and that this part of expansion does not give a proper result.
To see this it is sufficient to observe that this naive term (as well as
other terms that involve Taylor operators) happen to involve infrared
divergences starting from some minimal order of the expansion.
On the other hand, the first factors of the form
$F_{\Gamma / \gamma} \circ {\cal P}$ involve ultraviolet divergences after
the degree of the insertion polynomial is enough great.
A non-trivial point is that these induced divergences are mutually
cancelled --- see below.
Appearance of these spurious divergences can be considered as the price
to be paid to have explicit and simplest formulae of the expansion.

Note than expansion (\ref{F2}) is in powers of the expansion parameter once
we keep the regularization.
If the initial diagram is UV and IR finite it is nevertheless worthwhile
to introduce regularization to have simple and explicit formulae
for the expansions.
In the limit $\ep \to 0$ one obtains
an expansion in powers and logs, the maximal power of the logarithm
being no greater than the number of loops. This is essentially
a property of asymptotic expansions without external
legs on a mass shell.

\vspace{3mm}

{\bf 3.} {\em  One-loop example. }

\vspace{1mm}

Expansion (\ref{F2}) holds as well for ultravioletly divergent diagrams,
provided one does not switch off the regularization.

Let us consider the simplest example: a one-loop propagator-type diagram
with a mass $m_1=m \neq0$ and $m_2 =0$.
The corresponding Feynman integral\footnote{In this example we
consider Euclidean space for simplicity.}
 is written as
\be
F_{\Gamma} (Q,m;\ep) = \frac{1}{(2\pi)^d}
\int \frac{\dd k} {(k^{2} +m^{2})(Q-k)^{2}} .
\label{F7}
\ee
In the limit $k^{2} \rightarrow \infty$ relevant
subgraphs are $\Gamma$, $\{ 1 \}$, and $\{ 2 \}$.
The subgraph $\{ 1 \}$ does not  contribute because it
generates a massless vacuum diagram which is zero in dimensional
regularization.

In accordance with (\ref{F2}), the contribution from $\gm = \Gm$
is given by (\ref{F6}) and looks like
\bea
{\cal T}_m F_{\Gamma} (Q,m;\ep) = \frac{1}{(2\pi)^d}
\int \frac{\dd k} {(Q-k)^{2}} {\cal T}_m \frac{1}{k^{2} +m^{2}}  \nn \\
= \frac{1}{(2\pi)^d}
\int \frac{\dd k}{(Q-k)^{2}}
\left\{ \frac{1}{k^{2}}  - \frac{m^2}{k^{4}}
+ \frac{m^4}{k^{6}} -\ldots \right\}  .
\label{F8}
\eea

The contribution from $\gm = \{2\}$ is
\be
F_{\Gamma / \gamma} \circ
{\cal T}_{q^{\gm}} F_{\gamma}
= \frac{1}{(2\pi)^d}
\int \frac{\dd k}{k^{2} +m^{2}}
{\cal T}_{q^{\gm}} \frac{1}{(Q-k)^{2}} .
\label{F9}
\ee
The (only) small external momentum $q^{\gm}$ for $\gm = \{2\}$ is
just $k$ --- the loop momentum of the whole graph. Applying once again
the formula for geometrical series
\be
{\cal T}_{k} \frac{1}{(Q-k)^{2}}
= {\cal T}_{k} \frac{1}{Q^2 - (2Qk-k^2)} = \sum_{n=0}^{\infty}
(Q^2)^{-1-n} (2Qk-k^2)^n ,
\label{F10}
\ee
using one loop formula, resp., for massless propagator type and vacuum massive
Feynman integrals, and summing up two contributions from the subgraphs
involved
one gets
\bea
F_{\Gm} (Q, m;\ep)
\; \stackrel{\mbox{\footnotesize$Q \to \infty$}}{\mbox{\Large$\sim$}} \;
= \sum_{\gamma} F_{\Gamma / \gamma}
\circ {\cal T}_{\ldots} F_{\gamma} \nn \\
= \frac{1}{(4\pi)^{d/2}} (\mu^2 / Q^2)^{\ep} \left\{
\frac{\Gm^2 (1-\ep) \Gm(\ep)}{\Gm(2-2\ep)} \;
_1 F_0 (2\ep-1; -m^2 / Q^2) \nn \right.\\
\left. +
(m^2 / Q^2) ^{1-\ep} \Gm(\ep-1)
\; _2 F_1 (1,\ep;2-\ep; -m^2 / Q^2) \right\} .
\label{F11}
\eea
For brevity, hypergeometrical functions are here used to represent
the series involved.

This diagram was used just an example of the general procedure
of writing down explicit results for asymptotic expansions.
Of course, result (\ref{F11}) can be obtained in a more efficient
way. Note, however, that at 2-loop level we shall use
the general result just for evaluation of diagrams --- see below.

\vspace{3mm}

{\bf 4.} {\em  The large momentum expansion of renormalized diagrams. }

\vspace{1mm}

In case one deals with renormalized diagrams the corresponding expansion
\cite{Gor,Che1,Che2,Che3} (proved in \cite{Smi1,Che3})
looks like
\be
RF_{\Gm} (\uQ /\rho, \uq, \um;\ep)
\; \stackrel{\mbox{\footnotesize$\uQ \to \infty$}}{\mbox{\Large$\sim$}} \;
\sum_{\gamma} \oR F_{\Gamma / \gamma} (\uq,\um;\ep)
\circ R {\cal T}_{\uq^{\gm}, \um^{\gm}}
F_{\gamma} (\uQ,\uq^{\gm}, \um^{\gm};\ep).
\label{F12}
\ee
In addition to (\ref{F2}) we use the following notation: $R$ is dimensional
renormalization (e.g. the MS-scheme), and $\oR$ is an
incomplete $R$-operation --- it does not include counterterms for subgraphs
of $\Gm / \gm$ that contain the vertex $v_{\gm}$ to which  the subgraph
$\gm$ was collapsed.

The renormalized expansion possesses all the above mentioned properties
of the unrenormalized expansion. For instance, it has both UV and IR
divergences that are mutually cancelled after making summation over
all the relevant subgraphs.

\vspace{3mm}

{\bf 5.} {\em  Explicitly finite expansion of renormalized diagrams. }

\vspace{1mm}

To see that spurious ultraviolet and infrared divergences are indeed
cancelled in (\ref{F2}) and (\ref{F12}) it is sufficient to write down
the expansions in a manifestly finite form. In \cite{Che3} it was proved
that (\ref{F12}) can be written as (see also
\cite{Che1,Che2,Smi1,Smi2})
\be
RF_{\Gm} (\uQ /\rho, \uq, \um;\ep)
\; \stackrel{\mbox{\footnotesize$\uQ \to \infty$}}{\mbox{\Large$\sim$}} \;
\sum_{\gamma}  F_{\Gamma / \gamma} (\uq,\um;\ep)
\circ R^* {\cal T}_{\uq^{\gm}, \um^{\gm}}
F_{\gamma} (\uQ,\uq^{\gm}, \um^{\gm};\ep).
\label{F13}
\ee
The changes are minimal: $\oR$ is replaced by $R$ and $R$ by $R^*$.
Here $R^*$ is the so-called $R^*$-operation introduced in
\cite{Che5} and developed in \cite{Che6,Che7}. It is a generalization
of the renormalization procedure for the case of IR divergences.
It is constructed using similarity of UV and IR divergences.
The basic
property of the $R^*$-operation is that it removes both UV and IR
divergences from arbitrary Feynman integral (without restriction to a
mass shell). This theorem was proved in \cite{Che7,Che3} (see also
\cite{Smi2}).

The $R^*$-operation originated as a powerful technique in
renormalization group calculations: for example it was successfully
applied in the world record 5-loop calculations of the beta function
and anomalous dimensions in the $\phi^4$-theory \cite{Che8}.

The manifestly finite expansion (\ref{F13}) was first found \cite{Che88}
for a particular example associated with the operator
product expansion and then justified for the general case of expansion
with one large momentum in \cite{Che2}.

\vspace{3mm}

{\bf 6.} {\em The large mass expansion.}

\vspace{1mm}

Explicit asymptotic expansions in other limits are written almost
in the same form as above expansions in the large momentum limit.
The only distinction is that, for each limit, it is necessary to
write down the sum over specific class of subgraphs. For example,
in the large mass limit when some masses are much greater than
the other masses and all the momenta, one has
the following analog of (\ref{F2}):
\be
F_{\Gm} (\uq, \uM,\um;\ep)
\; \stackrel{\mbox{\footnotesize$\uM \to \infty$}}{\mbox{\Large$\sim$}} \;
\sum_{\gamma} F_{\Gamma / \gamma} (\uq,\um;\ep)
\circ {\cal T}_{\uq^{\gm}, \um^{\gm}}
F_{\gamma} (\uq^{\gm}, \uM,\um^{\gm};\ep),
\label{F14}
\ee
where the sum is now over subgraphs $\gamma$ of $\Gamma$ such that each
$\gamma$ (it may be disconnected)
(a) contains all the lines with the large mass
momenta and (b) consists of connectivity components that are
1PI with respect to lines with small masses.

For example, if all the masses of the diagram are large, then the
combinatorics
of the expansion is trivial because only the graph itself contributes to the
sum so that the expansion consists only of the naive part and
is just Taylor expansion in small momenta and masses:
\be
F_{\Gm} (\uq, \uM;\ep)
\; \stackrel{\mbox{\footnotesize$\uM \to \infty$}}{\mbox{\Large$\sim$}} \;
{\cal T}_{\uq}
F_{\Gm} (\uq, \uM;\ep).
\label{F15}
\ee
Here of course one can put $\ep=0$ if the initial diagram is convergent.

\vspace{3mm}

{\bf 7.} {\em Evaluation of master 2-loop diagram:
pure large mass expansion.}

\vspace{1mm}

We shall see later how general theorems on asymptotic expansions
in momenta and masses
can
be successfully applied for calculation of diagrams. But let us first
consider an example of combinatorially trivial expansion (\ref{F15})
when the knowledge of general results is unnecessary.
Let us consider the master
two-loop diagram with non-zero masses (see Fig.~1a)
in the large mass limit.
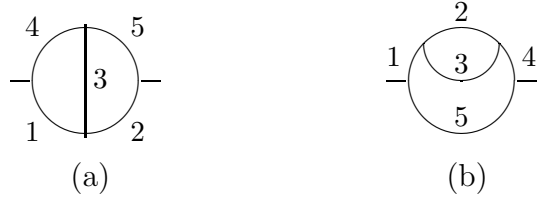
\begin{figure}[htb]
\begin{picture}(150,30)(-10,0)
\put (40,15) {\circle{15}}
\put (40,7.5) {\line(0,1){15}}
\put (30,15) {\line(1,0){2.5}}
\put (47.5,15) {\line(1,0){2.5}}
\put (32,7) {{\small 1}}
\put (46,7) {{\small 2}}
\put (32,21) {{\small 4}}
\put (46,21) {{\small 5}}
\put (41,14) {{\small 3}}
\put (38,1) {(a)}
\put (90,15) {\circle{15}}
\put (90,20) {\oval(10,10)[b]}
\put (80,15) {\line(1,0){2.5}}
\put (97.5,15) {\line(1,0){2.5}}
\put (80,17) {{\small 1}}
\put (98,17) {{\small 4}}
\put (89,16) {{\small 3}}
\put (89,23) {{\small 2}}
\put (89,9) {{\small 5}}
\put (88,1) {(b)}
\end{picture}
\caption{Two-loop self-energy diagrams.}
\end{figure}

The corresponding analytically and dimensionally regularized
Feynman integral up to a trivial factor is written as
\be
\label{defJ}
J\left( \unu ; \um ; q \right)
= \int \int
\frac{\dd^d k \; \dd^d l}
      {D_1^{\nu_1} D_2^{\nu_2} D_3^{\nu_3} D_4^{\nu_4} D_5^{\nu_5}} ,
\ee
where $\nu_i$ are the powers
of the denominators $D_i \equiv p_i^2 - m_i^2 + i0$, $\; p_i$ being
the momentum of the corresponding line ($p_i$ are constructed from
the loop momenta $k$ and $l$ and the external momentum $q$).
This is a very complicated function of five dimensionless variables.
The most interesting case is when all $\nu$'s are integer.
The cases when some of the $\nu$'s are zero correspond to reducing
lines in Fig.~1a to points. In such a way,
self-energy diagrams with four or three internal lines can
also be described. Moreover, by trivial decomposition (partial
fractioning) of the first and the fourth denominators (provided
that $\nu_1$ and $\nu_4$ are integer) one can reduce the integral
corresponding to Fig.~1b to the integrals (\ref{defJ}) with
$\nu_1$ or $\nu_4$ equal to zero (such a decomposition is required
only if $m_1 \neq m_4$). So, in the general case of self-energy
diagrams (with integer $\nu$'s) it is sufficient to consider only
the integrals (\ref{defJ}).

In \cite{DT} a large number of first terms of the large mass expansion
was analytically calculated. The problem was reduced to
calculation of vacuum two-loop Feynman integrals with three different masses:
\be
\label{defI}
I\left( \unu ; \um \right) =
\int \int
\frac{\mbox{d}^d k \; \mbox{d}^d l}
   {[p^2-m_1^2]^{\nu_1} [q^2-m_2^2]^{\nu_2}
    [(p-q)^2-m_3^2]^{\nu_3}} .
\ee
For example, in the case of three different masses and unique indices
$\nu_i=1$ one has \cite{DT}
\bea
I\left(1,1,1 ; \um \right)
= \pi^{4-2\ep} (m_3^2) \frac{A(\ep)}{2} \left\{
- \frac{1}{\ep^2} (1+x+y)
+ \frac{2}{\ep} (x \ln x +y \ln y) \right. \nn \\
\left. - x \ln^2 x - y \ln^2 y +
(1-x-y) \ln x \ln y - \lm^2 \Phi (x,y) \right\} ,
\label{F16}
\eea
with
\bea
x= m_1^2 / m_3^2, \; y= m_2^2 / m_3^2, \nn \\
\lm = \sqrt{(1-x-y)^2-4xy},\;\; A(\ep) = \Gm^2(1+\ep) / (1-\ep)(1-2\ep),\nn \\
\Phi(x,y) = \frac{1}{\lm} \left\{
2 \ln ((1+x-y-\lm)/2) \ln ((1-x+y-\lm)/2) -\ln x \ln y \right. \nn \\
\left. -2 \mbox{Li}_2 ((1+x-y-\lm)/2) -2 \mbox{Li}_2 ((1-x+y-\lm)/2)
+ \pi^2/3 \right\},
\label{F17}
\eea
and Li$_2 (z)$ di-logarithm.
(Similar results were obtained in \cite{vdB}.)

By comparing the obtained results with numerical calculations based
on a two-fold parametric representation \cite{Kre}, in \cite{DT} it
was shown that several first terms of the expansion
provide a very good approximation for the diagram if the value of the
momentum is not too close to a threshold.

\vspace{3mm}

{\bf 8.} {\em Evaluation of master 2-loop diagram:
the large momentum expansion.}

\vspace{1mm}

For the large momentum limit, the naive Taylor expansion alone does not
produce a correct result so that we need to apply general formula (\ref{F2}).
The corresponding set of relevant subgraphs is shown in Fig.~2.
\begin{figure}[htb]
\begin{picture}(150,100)(-25,0)
\put (0,89) {\small Type 1:}
\put (30,90) {\circle{10}}
\put (25,90) {\circle*{0.8}}
\put (35,90) {\circle*{0.8}}
\put (30,95) {\circle*{0.8}}
\put (30,85) {\circle*{0.8}}
\put (30,85) {\line(0,1){10}}
\put (0,69) {\small Type 2:}
\put (30,70) {\oval(10,10)[t]}
\put (30,70) {\oval(10,10)[br]}
\put (25,70) {\circle*{0.8}}
\put (35,70) {\circle*{0.8}}
\put (30,75) {\circle*{0.8}}
\put (30,65) {\circle*{0.8}}
\put (30,65) {\line(0,1){10}}
\put (50,70) {\oval(10,10)[t]}
\put (50,70) {\oval(10,10)[bl]}
\put (45,70) {\circle*{0.8}}
\put (55,70) {\circle*{0.8}}
\put (50,75) {\circle*{0.8}}
\put (50,65) {\circle*{0.8}}
\put (50,65) {\line(0,1){10}}
\put (70,70) {\oval(10,10)[b]}
\put (70,70) {\oval(10,10)[tr]}
\put (65,70) {\circle*{0.8}}
\put (75,70) {\circle*{0.8}}
\put (70,75) {\circle*{0.8}}
\put (70,65) {\circle*{0.8}}
\put (70,65) {\line(0,1){10}}
\put (90,70) {\oval(10,10)[b]}
\put (90,70) {\oval(10,10)[tl]}
\put (85,70) {\circle*{0.8}}
\put (95,70) {\circle*{0.8}}
\put (90,75) {\circle*{0.8}}
\put (90,65) {\circle*{0.8}}
\put (90,65) {\line(0,1){10}}
\put (0,49) {\small Type 3:}
\put (30,50) {\circle{10}}
\put (25,50) {\circle*{0.8}}
\put (35,50) {\circle*{0.8}}
\put (30,55) {\circle*{0.8}}
\put (30,45) {\circle*{0.8}}
\put (0,29) {\small Type 4:}
\put (30,30) {\oval(10,10)[tl]}
\put (30,30) {\oval(10,10)[br]}
\put (30,25) {\line(0,1){10}}
\put (25,30) {\circle*{0.8}}
\put (35,30) {\circle*{0.8}}
\put (30,35) {\circle*{0.8}}
\put (30,25) {\circle*{0.8}}
\put (50,30) {\oval(10,10)[bl]}
\put (50,30) {\oval(10,10)[tr]}
\put (50,25) {\line(0,1){10}}
\put (45,30) {\circle*{0.8}}
\put (55,30) {\circle*{0.8}}
\put (50,35) {\circle*{0.8}}
\put (50,25) {\circle*{0.8}}
\put (0,9) {\small Type 5:}
\put (30,10) {\oval(10,10)[t]}
\put (25,10) {\circle*{0.8}}
\put (35,10) {\circle*{0.8}}
\put (30,15) {\circle*{0.8}}
\put (50,10) {\oval(10,10)[b]}
\put (45,10) {\circle*{0.8}}
\put (55,10) {\circle*{0.8}}
\put (50,5) {\circle*{0.8}}
\end{picture}
\caption{The subgraphs contributing to the large momentum
expansion (2).}
\end{figure}
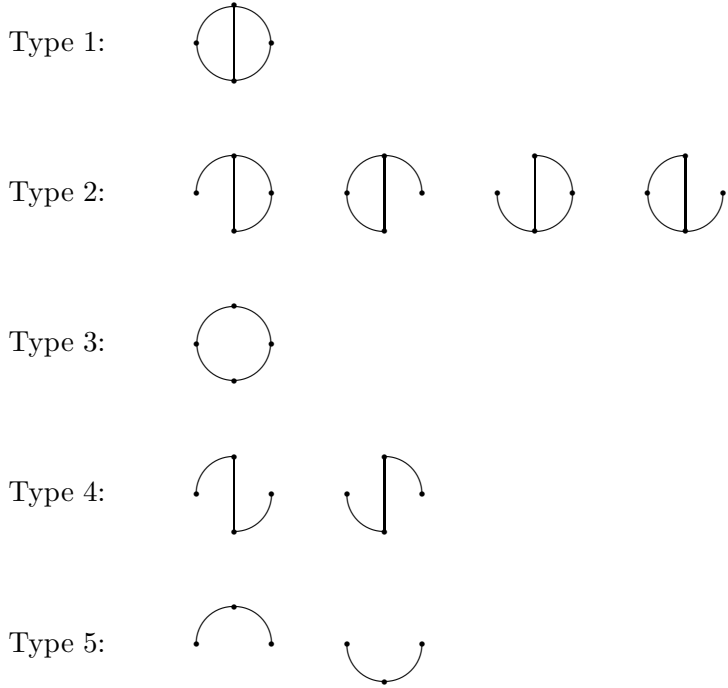

In  case $\gamma=\Gamma$ (Type 1), all denominators of (\ref{defJ})
should be expanded in masses:
\be
\label{1term}
{\cal{T}}_{\um} \; J(\unu ; \um ; Q)
= \sum_{j_1, \ldots , j_5 = 0}^{\infty}
\frac{(\nu_1)_{j_1} \ldots (\nu_5)_{j_5} }{j_1 ! \ldots j_5 !}
(m_1)^{j_1} \ldots (m_5)^{j_5} \;
J(\unu + \underline{j} ; \underline{0} ; Q) ,
\ee
where
\be
\label{poch}
(\nu)_j \equiv \frac{\Gamma(\nu+j)}{\Gamma(\nu)}
\ee
is the Pochhammer symbol .

Note that if we consider the case $\nu_1= \ldots = \nu_5 =1$, the first
term of the expansion (\ref{1term}) (with $j_1 = \ldots = j_5 = 0$)
gives the well-known result \cite{Ros,Geg}:
$ - 6 \zeta(3) \pi^4/Q^2 $. Two-loop massless integrals with higher integer
powers of denominators occurring on the right-hand side
of (\ref{1term}) can be
evaluated by use of the integration-by-parts technique
\cite{IBP} or by Gegenbauer polynomial technique \cite{Geg}.

A typical contribution of the second type (see
Fig.~2), when $\gamma$ is obtained from $\Gamma$ by removing line 1,
looks like
\be
\label{2term}
\! \int \! \frac{\mbox{d}^d k}{{[k^2 \! - \! m_1^2]}^{\nu_1}} \;
{\cal{T}}_{m_2, \ldots , m_5; k}
\int \! \frac{\mbox{d}^d l}
   {{[l^2 \! - \! m_2^2]}^{\nu_2} {[(k\!-\!l)^2\! -\! m_3^2]}^{\nu_3}
     {[(Q\!-\!k)^2 \!- \!m_4^2]}^{\nu_4}  {[(Q\!-\!l)^2\! - \!m_5^2]}^{\nu_5}
}
 .
\ee
After expanding the integrand of the $l$-integral in masses and $k$, one
obtain products of massless one-loop integrals and
massive tadpoles with numerators \cite{DST}.

For Type 3 one obtains:
\bea
\label{3term}
\! \int \! \frac{\mbox{d}^d k}{{[k^2 \! - \! m_3^2]}^{\nu_3}} \;
\hspace{12cm} \nonumber \\ \times
{\cal{T}}_{m_1, m_2, m_4 , m_5; k}
\int \! \frac{\mbox{d}^d l}
   {{[(Q\!+\! k \! - \!l)^2 \! - \! m_1^2]}^{\nu_1}
    {[(Q\!-\!l)^2\! -\! m_2^2]}^{\nu_2}
     {[(k\!-\!l)^2 \!- \!m_4^2]}^{\nu_4}  {[l^2\! - \!m_5^2]}^{\nu_5} } .
\eea
The resulting  integrals are of the same type as in the previous case.

For Type 4 there are no loop integrations in the subgraph $\gamma$, and we get
for the first contribution of the fourth type:
\bea
\label{4term}
\! \int \! \int \! \frac{\mbox{d}^d k}{{[k^2 \! - \! m_1^2]}^{\nu_1}}
                   \frac{\mbox{d}^d l}{{[l^2 \! - \! m_5^2]}^{\nu_5}} \;
\hspace{9cm} \nonumber \\ \times
{\cal{T}}_{m_2, m_3, m_4 ; k, l}
\left( \frac{1}{ {[(Q\!-\!l)^2\! -\! m_2^2]}^{\nu_2}
 {[(Q\! -\!k\!-\!l)^2\! -\! m_3^2]}^{\nu_3}
 {[(Q\!-\!k)^2 \!- \!m_4^2]}^{\nu_4} } \right) .
\eea
As a result, one obtains products of two one-loop tadpoles with numerators
(also for the second contribution).

For Type 5 one obtains non-trivial two-loop vacuum
integrals. For example, the first contribution of the fifth type
gives:
\bea
\label{5term}
 \int \! \int \! \frac{\mbox{d}^d k \; \mbox{d}^d l}
          {{[k^2 \! - \! m_1^2]}^{\nu_1} {[l^2 \! - \! m_2^2]}^{\nu_2}
               {[(k-l)^2 \! - \! m_3^2]}^{\nu_3}}
\hspace{5cm} \nonumber \\ \times
{\cal{T}}_{m_4, m_5; k, l}
\left( \frac{1} {{[(Q\!-\!k)^2 \!- \!m_4^2]}^{\nu_4}
        {[(Q\!-\!l)^2\! -\! m_5^2]}^{\nu_5}} \right) .
\eea
Expanding the denominators, one obtains two-loop vacuum integrals with
numerators, which are then reduced \cite{DST} to the same
integrals (\ref{defI}) as in the opposite limit.

For example, in the case $\nu_i =1, i=1,\ldots,5$ and $\ep=0$ one has
\cite{DST}
\be
\label{defMj}
\left. J(1, \ldots ,1; \um; Q) \right|_{d=4}
=  J( \um; Q)
= 
- \frac{\pi^4}{Q^2}
{\cal M}(\um; Q)
=  - \frac{\pi^4}{Q^2}
\sum_{j=0}^{\infty} \frac{{\cal M}_j}{(Q^2)^j} , 
\ee
where the coefficient functions ${\cal M}_j$ include powers
of masses and logarithms of masses and momentum squared.
It is easy to see that the only integral contributing to
${\cal M}_0$ is $J^{(0)}(1,1,1,1,1)$  in
(\ref{1term})
and that the expansion starts from
\be
\label{M0}
{\cal M}_0 = 6 \zeta(3) .
\ee

The ${\cal M}_1$ term already includes contributions of all
terms (\ref{1term})--(\ref{5term}) (with the exception of
(\ref{4term}) that begins to contribute starting from
${\cal M}_2$); this yields
\bea
\label{M1}
{\cal M}_1 &=& \frac{m_1^2}{2}
\left\{ \ln^2 \left( -\frac{Q^2}{m_1^2} \right)
+ 4 \ln \left( -\frac{Q^2}{m_1^2} \right)
- \ln \frac{m_2^2}{m_1^2} \; \ln \frac{m_3^2}{m_1^2} + 6
\right\}
\nonumber \\[2mm]
&&+ \left\{ \mbox{analogous terms with} \; m_2^2, \; m_4^2, \; m_5^2 \right\}
\nonumber \\[2mm]
&&+ \frac{m_3^2}{2}
\left\{ 2 \ln^2 \left( -\frac{Q^2}{m_3^2} \right)
+ 4 \ln \left( -\frac{Q^2}{m_3^2} \right)
- \ln \frac{m_1^2}{m_3^2} \; \ln \frac{m_2^2}{m_3^2}
- \ln \frac{m_4^2}{m_3^2} \; \ln \frac{m_5^2}{m_3^2}
\right\}
\nonumber \\[2mm]
&&+ \frac{1}{2} \left\{ F (m_1^2, m_2^2, m_3^2) + F (m_4^2, m_5^2, m_3^2)
\right\} \, ,
\eea
where the symmetric function $F$ is defined by
\be
\label{defF}
F (m_1^2, m_2^2, m_3^2) \equiv m_3^2
\lambda^2  \left( x, y \right)
\Phi \left( x, y \right)
\ee
and (\ref{F17}). By use of the
{\sf REDUCE} system \cite{RED}, analytical results for
the general massive case of the
integral (\ref{defMj}) (when all five masses are arbitrary)
were obtained
for the coefficient functions up to ${\cal M}_6$ \cite{DST}.

These results are in complete agreement with known explicit
expressions in  cases where non-zero masses are equal
\cite{Bro1,Bro2,ST}.
Furthermore, as in the large mass limit, a comparison with numerical
calculations based on the same two-fold representation \cite{Kre}
shows that only first few terms provide a very good
approximation provided we are not close to the highest threshold.

Thus, in situations when one is below the lowest threshold or above
highest threshold one can substitute the general master
diagram by sufficient number of terms of small (resp., large)
momentum expansion.

\vspace{3mm}

{\bf 9.} {\em Evaluation of master 2-loop diagram:
the large mass expansion with zero \linebreak thresholds.}

\vspace{1mm}

Let us now consider the small momentum (large mass) expansion when
some masses are large while the other masses are zero.
If there no zero thresholds one can apply the large mass expansion
of Section 7.
There exist four independent zero-threshold
configurations
\footnote{2PT and 3PT mean two- and
three-particle thresholds, respectively.}:

Case 1: one zero-2PT (e.g. $m_2 = m_5 = 0$);

Case 2: two zero-2PT's ($m_1 = m_2 = m_4 = m_5 = 0$);

Case 3: one zero-3PT (e.g. $m_2 = m_3 =m_4 = 0$);

Case 4: one zero-2PT and one zero-3PT (e.g. $m_2 = m_3 = m_4 = m_5 = 0$).

Note that, in case 1, one more vanishing mass (e.g. $m_3$
 or $m_4$) does not produce a new zero-threshold configuration,
and the corresponding cases can be considered together with case 1,
e.g.:

Case 1a $=$ Case 1 with $m_3 = 0$;

Case 1b $=$ Case 1 with one more mass (not $m_3$) equal to zero.

The set of relevant subgraphs that should be involved in general
formula (\ref{F14}) is shown in Fig.~3.
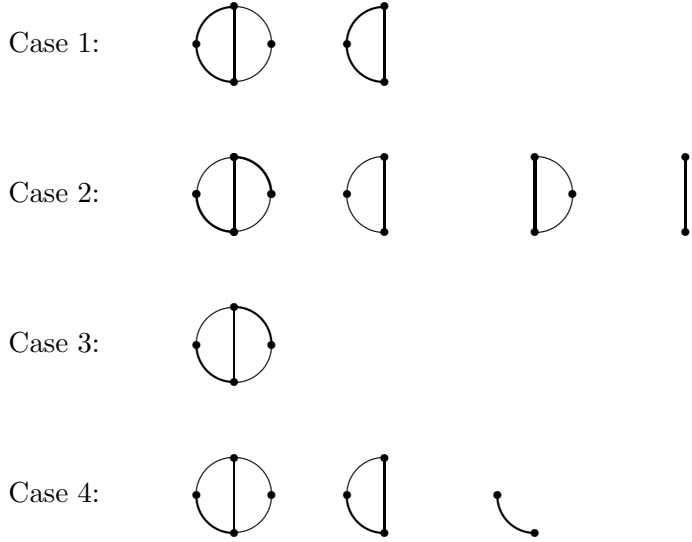
\begin{figure}[htb]
\begin{picture}(150,80)(-25,0)
\put (0,69) {\small Case 1:}
\thicklines
\put (30,70) {\oval(10,10)[tl]}
\put (30,70) {\oval(10,10)[bl]}
\put (30,65) {\line(0,1){10}}
\thinlines
\put (30,70) {\oval(10,10)[tr]}
\put (30,70) {\oval(10,10)[br]}
\put (25,70) {\circle*{1}}
\put (35,70) {\circle*{1}}
\put (30,75) {\circle*{1}}
\put (30,65) {\circle*{1}}
\thicklines
\put (50,70) {\oval(10,10)[tl]}
\put (50,70) {\oval(10,10)[bl]}
\put (50,65) {\line(0,1){10}}
\thinlines
\put (45,70) {\circle*{1}}
\put (50,75) {\circle*{1}}
\put (50,65) {\circle*{1}}
\put (0,49) {\small Case 2:}
\put (30,50) {\circle{10}}
\thicklines
\put (30,45) {\line(0,1){10}}
\thinlines
\put (25,50) {\circle*{1}}
\put (35,50) {\circle*{1}}
\put (30,55) {\circle*{1}}
\put (30,45) {\circle*{1}}
\put (50,50) {\oval(10,10)[tl]}
\put (50,50) {\oval(10,10)[bl]}
\thicklines
\put (50,45) {\line(0,1){10}}
\thinlines
\put (45,50) {\circle*{1}}
\put (50,55) {\circle*{1}}
\put (50,45) {\circle*{1}}
\put (70,50) {\oval(10,10)[tr]}
\put (70,50) {\oval(10,10)[br]}
\thicklines
\put (70,45) {\line(0,1){10}}
\thinlines
\put (75,50) {\circle*{1}}
\put (70,55) {\circle*{1}}
\put (70,45) {\circle*{1}}
\thicklines
\put (90,45) {\line(0,1){10}}
\thinlines
\put (90,55) {\circle*{1}}
\put (90,45) {\circle*{1}}
\put (0,29) {\small Case 3:}
\put (30,30) {\oval(10,10)[tl]}
\put (30,30) {\oval(10,10)[br]}
\put (30,25) {\line(0,1){10}}
\thicklines
\put (30,30) {\oval(10,10)[tr]}
\put (30,30) {\oval(10,10)[bl]}
\thinlines
\put (25,30) {\circle*{1}}
\put (35,30) {\circle*{1}}
\put (30,35) {\circle*{1}}
\put (30,25) {\circle*{1}}
\thicklines
\put (30,50) {\oval(10,10)[tr]}
\put (30,50) {\oval(10,10)[bl]}
\thinlines
\put (25,50) {\circle*{1}}
\put (35,50) {\circle*{1}}
\put (30,55) {\circle*{1}}
\put (30,45) {\circle*{1}}
\put (0,9) {\small Case 4:}
\put (30,10) {\oval(10,10)[t]}
\put (30,10) {\oval(10,10)[br]}
\put (30,5) {\line(0,1){10}}
\thicklines
\put (30,10) {\oval(10,10)[bl]}
\thinlines
\put (25,10) {\circle*{1}}
\put (35,10) {\circle*{1}}
\put (30,15) {\circle*{1}}
\put (30,5) {\circle*{1}}
\put (50,10) {\oval(10,10)[tl]}
\put (50,5) {\line(0,1){10}}
\thicklines
\put (50,10) {\oval(10,10)[bl]}
\thinlines
\put (45,10) {\circle*{1}}
\put (50,15) {\circle*{1}}
\put (50,5) {\circle*{1}}
\thicklines
\put (70,10) {\oval(10,10)[bl]}
\thinlines
\put (65,10) {\circle*{1}}
\put (70,5) {\circle*{1}}
\end{picture}
\caption{The subgraphs contributing to the large mass
expansions (14) with zero thresholds. Thick and thin lines
correspond respectively to massive and massless particles.}
\end{figure}

In the resulting expansion one has \cite{BDST}

(a) two-loop vacuum diagrams with two (or one) massive and
    one (or two) massless lines;

(b) products of a one-loop massless diagram with external
    momentum $k$ and a one-loop massive tadpole;

(c) two-loop massless diagrams with one (or two) powers of
    the denominators being non-positive.

The results of calculation are described in \cite{BDST}.
In particular, for unique indices one has
\be
\label{J-exp}
J(\um; q)
\equiv J(1,1,1,1,1;\um; q) =
- \pi^4 \sum_{j=0}^{\infty} C_j \; (q^2)^j ,
\ee
where, for zero-threshold configurations, the coefficients $C_j$ can
depend on logarithms of $q^2$:
\be
\label{C_j}
C_j = C_j^{(2)} \ln^2(-q^2) + C_j^{(1)} \ln(-q^2) + C_j^{(0)}.
\ee

For cases when one has only one massive parameter
$m$ (all non-zero masses are equal),
the results obtained are in a full agreement with \cite{Bro1,Bro2,ST}.

For Case 1 with three non-equal masses
one has, in particular,
\bea
\label{C0-case1-3m}
C_0 =
\frac{1}{4 (m_1^2 - m_3^2)(m_4^2 - m_3^2) (m_1^2 - m_4^2)}
\left\{ 2 \left(\ln\left(-\frac{q^2}{m_1^2}\right)
              + \ln\left(-\frac{q^2}{m_4^2}\right) - 4 \right)
\right.
\nonumber \\
\times
        \left( m_1^2 (m_4^2 - m_3^2) \ln\frac{m_1^2}{m_3^2}
             - m_4^2 (m_1^2 - m_3^2) \ln\frac{m_4^2}{m_3^2} \right)
\nonumber \\
       + (m_1^2 - m_3^2) (m_4^2 - m_3^2)
          \ln\frac{m_1^2}{m_4^2}
  \left( \ln\frac{m_1^2}{m_3^2} + \ln\frac{m_4^2}{m_3^2} \right)
\nonumber \\
       - 2 m_3^2 (m_1^2 - m_4^2)
          \ln\frac{m_1^2}{m_3^2} \ln\frac{m_4^2}{m_3^2}
       + 2 (m_1^2 + m_3^2) (m_4^2 - m_3^2) {\cal{H}}(m_1^2, m_3^2)
\nonumber \\
\left.
  - 2 (m_4^2 + m_3^2) (m_1^2 - m_3^2) {\cal{H}}(m_4^2, m_3^2)
\frac{}{} \right\} ,
\eea
where ${\cal{H}}$ is a dimensionless function
defined as
\be
\label{defH}
{\cal{H}}(m_1^2, m_2^2) = 2\Li{2}{1- \frac{m_1^2}{m_2^2}}
               + \frac{1}{2} \ln^2\left(\frac{m_1^2}{m_2^2}\right).
\ee
In \cite{BDST} the coefficients were analytically calculated up to
$C_3$.

In Case 3 with two different non-zero masses, one has, for instance,
\be
\label{C0-case3-2m}
C_0 =
\frac{1}{4 m_1^2 m_5^2}
\left\{ - (m_1^2 + m_5^2)
       \left( \ln^2{\frac{m_1^2}{m_5^2}} + \frac{2 \pi^2}{3} \right)
        - 2 (m_1^2 - m_5^2)\; {\cal{H}}(m_1^2, m_5^2)
\right\} ,
\ee
\bea
\label{C1-case3-2m}
C_1 =
\frac{1}{8 m_1^4 m_5^4 (m_1^2 - m_5^2)}
\left\{ 2 m_1^2 m_5^2 (m_1^2 - m_5^2)
        \left( \ln\left(-\frac{q^2}{m_1^2}\right)
            + \ln\left(-\frac{q^2}{m_5^2}\right) - 3 \right)
\right.
\nonumber \\[2mm]
- (m_1^2 - m_5^2) (m_1^4 + m_5^4)
  \left( \ln^2{\frac{m_1^2}{m_5^2}} + \frac{2 \pi^2}{3} \right)
+ 2 m_1^2 m_5^2 (m_1^2 + m_5^2) \ln{\frac{m_1^2}{m_5^2}}
\nonumber \\[2mm]
\left. \frac{}{}
- 2 (m_1^2 - m_5^2)^2 \; (m_1^2 + m_5^2) \; {\cal{H}}(m_1^2, m_5^2)
\right\}  ,
\eea
In \cite{BDST} these coefficients were analytically calculated up to
$C_5$.

By comparison with numerical integration \cite{Kre,Fuj,Kre3}
it was found that
substitution of the initial function by several terms
of its asymptotic expansion provides a satisfactory agreement
unless we are very close to the first non-zero threshold.

\vspace{3mm}

{\bf 10.} {\em Other approaches and possibilities.}

\vspace{1mm}
The general formulae of asymptotic expansions were applied for
calculation of Feynman diagrams at 3- and 4-loop level in other
various situations ---
see e.g. \cite{Che9}.

Results for other two-loop massive diagrams, e.g.
the `setting sun' ($\equiv$ London transport) diagram, as well as
presentation of other methods can be found in
\cite{Fuj,Boe,Buza,BT,Ghi,FT,Lun}.
It looks natural to apply general results for asymptotic expansions
in momenta and masses for massive 3-point 2-loop diagrams. In
`combinatorially trivial' case of the large mass expansion
calculations were performed in \cite{FT}.
For 3-point diagrams one can also use analogous two-fold
parametric representation \cite{Kre3} for comparison.

Note that there are no general results for limits with some
legs at a mass shell in specifically Minkowskian situations.
See, e.g., \cite{SL} for discussion of asymptotic behavior in the
Sudakov limit.

\vspace{5mm}

{\em Acknowledgements.}
I am grateful to the organizers of the Ringberg Workshop
for kind hospitality. I am much indebted to K.G.~Chetyrkin
and A.I.~Davydychev for bibliographical comments
and careful reading of the manuscript.

\end{document}